\documentclass{article}
\usepackage{amsfonts}
\usepackage{amsmath}

\input{tcilatex}
\begin{document}

\title{Thermodynamic Treatment of High Energy Heavy Ion Collision}
\author{Wedad AL-Harbi$^1$ and Tarek Hussein $^2$ \\
%EndAName
$^{1}$ Physics Department- Sciences Faculty For girls,\\
King Abdulaziz University\\
$^{2}$ Physics Department, Faculty of Science, \\
Cairo University, 12613 Giza, Egypt }
\maketitle

\begin{abstract}
The hadron production in heavy ion collision is treated in the framework of
thermodynamic vision. Thermodynamic system formed during central collision
of Pb-Pb at high energies is considered, through which binary collision is
assumed among the valance quarks. The partition function of the system is
calculated; accordingly the free available energy, the entropy and the
chemical potential are calculated. The concept of string fragmentation and
defragmentation are used to form the newly produced particles. The average
multiplicity of the newly produced particles are calculated and compared
with the recent experimental results.\newline
\end{abstract}

\section{Introduction}

In a wide energy range, there are some common aspects of heavy ion reaction
dynamics. The energies are large enough and the masses of ions are also
large, to consider the heavy ions as classical particles. Their De Broglie
wavelength is much less than the typical nuclear sizes. Quantum effects
influence the underlying microscopic dynamics only, which can be included in
the equation of state, in the transport coefficients, or in the kinetic
theory describing the reactions. During the collision and assuming straight
line trajectories there can be target and projectile spectators in a
collision \textbf{[1-3]} The rest of the nucleons may hit each other on the
way forming a participant zone with both target and projectile nucleons in
it. The most interesting phenomena and the new physics are in the
participant zone. The spectator regions also provide us with interesting
phenomena. Realistically, the nucleons do not propagate along exactly
straight trajectories. Deviations from straight propagations are observed
even at the highest energies of 200 GeV per nucleon. According to the fluid
dynamical model, considerable collective sideward motion is generated 
\textbf{[4]}. We shall deal with the collision problem using statistical
physics concepts according to the following vision:

A heavy ion reaction is a dynamical system of a few hundred nucleons. This
is a large number but still far from the continuum, so that deviations from
infinite matter limit are important. On the other hand, the number of
particles participating in a reaction is large enough that the signs of
collective matter like behavior can be clearly observed. This is an
interesting territory in statistical physics of small but collective
systems. The methods developed in this field are unique and may also be
applicable in other \textquotedblleft small\textquotedblright\ statistical
systems.

However when Quark Gluon Plasma is formed the number of quanta increases to
a large extent\textbf{\ [5-6]}. The plasma can already be considered as a
continuum, and finite particle effects should be small.

On the other hand the heavy ion reaction is a rapid dynamical process. The
question of phase transitions in a dynamical system is still an open field
of research. Heavy ion physics may contribute to this field at two points:
i) the dynamics of the phase transitions in "small" systems, and ii) the
dynamics of the phase transitions in ultra-relativistic systems where the
energy of the system is much higher than the rest mass of the particles.

In this context, we shall deal with the problem from the statistical
thermodynamic point of view. In other words a hypothetical model will be
tailored according to the following concepts:

a.\qquad During the heavy ion collisions, the participant nucleons form a
thermodynamic system by a cylindrical cut of the projectile through the
target nucleus.

b.\qquad A fireball oriented trend is used considering valence quarks of the
nucleons as the constituent particles of the overlap region.

c.\qquad Binary collisions are assumed among the existing valence and the
created sea quarks. Accordingly a significant increase in energy density of
the system leads to an environment eligible to create more particles in a
frame of grand canonical ensemble.

d.\qquad A simple quark wave function is assumed to be used in calculating
the partition function of the system.

e.\qquad A power series expansion of the partition function is assumed and
we will consider only the first few terms that fit the boundary conditions
of the system.

According to the above mentioned assumptions we calculate the newly created
particle multiplicity produced in $Pb+Pb$ collisions to be compared with the
recently published experimental data. In doing this we should take into
consideration the following points:

i)\qquad The leading baryon is hardly stopped,

ii)\qquad In the rapidity region between the projectile and the target,
secondary charged particles (mesons $\pi ^{+}$; $\pi ^{-}$; $\pi ^{0}$; $%
\mathit{K}^{+};\mathit{K}^{-}$, etc.) are created through string mechanism 
\textbf{[7-8]}.

The paper is organized in four sections after this soft introduction. In
section 2 we represent the geometry and the formulation of the model.
Results and discussion are given in section 3. Eventually conclusive remarks
are given in section 4.

\bigskip

\section{Thermodynamic Treatment of the nuclear System}

In the hard sphere model, one effective radius $r=0.4fm$ \textbf{[9-10]} is
chosen for all quark collisions. Elastic and inelastic collisions are
considered which are supposed to be dominant at temperatures $T\approx
|120-200MeV.$ The calculations are done in the framework of the Boltzmann
equation with the Boltzmann statistical distribution functions and the real
gas equation of state \textbf{[1] , [11]}.

The Boltzmann equation is used to find the distribution function in phase
space of the thermodynamic system. It is a function in space coordinate,
momentum and time. In a system of large number of particles and a special
form of interaction potential, it will be difficult to get perfect full
solution of the equation. However many trials were done\textbf{\ [1] }to get
approximate solution to the Boltzmann equation in a form of conversing power
series. The first term represents the equilibrium state (zero order term).
The next terms represent the higher order corrections to describe the shift
due to the non-equilibrium state. These terms are time dependent and include
a time parameter that measure how far from the equilibrium states the
particles are produced. On putting the time parameter tends to zero, all the
higher terms of the series vanishes and the system approach equilibrium and
described only by the zero order term. Another trial to find phase space
distribution function in a pre-equilibrium state is the following:

Due to the none-equilibrium state the temperature is not homogeneous allover
the system, but instead there will be a temperature gradient with minimum
entropy. The approach is to divide the system into small stripes
(subsystems) each has its local equilibrium with a specific temperature and
local equilibrium distribution. The overall phase space distribution will be
the sum over the distributions of the subsystems. This is what we did in our
calculation.

The Boltzmann statistical approximation allows one to conduct precise
numerical calculations of transport coefficients in the hadron gas and to
obtain some relatively simple relativistic analytical closed-form
expressions.

For particles having spin, the differential cross sections were averaged
over the initial spin states and summed over the final ones.

The local equilibrium distribution functions are: 

\begin{subequations}
\begin{equation}
f_{k}^{0}=e^{(\mu _{k}-p_{k}^{\mu }U_{\mu })/T}  \tag{2-1}
\end{equation}

\bigskip Where, $%
%TCIMACRO{\U{3bc} }%
%BeginExpansion
\mu
%EndExpansion
k$ is the chemical potential of the kth particle species, $T$ is the
temperature and $U%
%TCIMACRO{\U{3bc} }%
%BeginExpansion
\mu
%EndExpansion
$ is the relativistic flow 4-velocity such that $U%
%TCIMACRO{\U{3bc} }%
%BeginExpansion
\mu
%EndExpansion
U^{%
%TCIMACRO{\U{3bc} }%
%BeginExpansion
\mu
%EndExpansion
}=1$ with frequently used consequence.

The distribution functions $f_{k}$ are found by solving the system of the
Boltzmann equations approximately with the form \textbf{\ [1]:}

\begin{equation}
f_{k}=f_{k}^{0}+f_{k}^{1}=f_{k}^{0}+f_{k}^{0}\varphi (x,p_{k})  \tag{2-2}
\end{equation}

\bigskip Because analytical expressions for the collision brackets are bulky
the Mathematica was used for symbolical and some numerical manipulations 
\textbf{\ [12]}. The numerical calculations are done also for temperatures
above $T=120MeV$.

Let us consider the case of the binary collisions between quarks of the
thermodynamic system that has been formed during the interaction of heavy
ions. The total Hamiltonian is:

\bigskip 
\begin{equation}
H_{tr}=\frac{1}{2M}\dsum\limits_{i-1}^{N}P_{i}^{2}+\sum_{i\neq
j}W_{ij}(r_{ij})  \tag{2-3}
\end{equation}

$p_{i}$ is the momentum of the ith quark and $W_{ij}$ is the binary
interaction potential energy among quarks$i\&j.$

The static quark potential at fixed spatial separation has been obtained
from an extrapolation of ratios of Wilson loops to infinite time separation.
As we have to work on still rather coarse lattices and need to know the
static quark potential at rather short distances (in lattice units) we have
to deal with violations of rotational symmetry in the potential. In our
analysis of the potential we take care of this by adopting a strategy used
successfully in the analysis of static quark potentials  \textbf{[13]} and
heavy quark free energies  \textbf{[14].} 

This procedure removes most of the short distance lattice artifacts. It
allows us to perform fits to the heavy quark potential with the 3-parameter
approach,

The quark-quark potential is given as 

\begin{equation}
W_{qq}(r)=-\frac{\alpha }{r}+\sigma r+c  \tag{2-4}
\end{equation}

The quark potential is graphically represented in \textbf{Fig. (1)}. It is
formed by 3 terms. The first is a Coulomb like; the second is string
repulsive potential that works in confinement the quarks inside the nucleon
bag. 

In this model, the number of created particles depends on the available
energy. Accordingly, we focus our calculation on getting information about
the relation between the energy and the number of interacting particles and
consequently the number of created particles 

The grand canonical ensemble considers large supersystem kept at constant $T$
and $P$ and consists of many subsystems that can exchange not only the
energy but also the number of particles. A number of particles and their
quantum numbers corresponding to their energy states specify a microstate in
the grand canonical ensemble. The particle abundance during the heavy ion
collision is much complicated and depends mainly on an environment in the
presence of catalyst necessary for the particle creation. The available
energy is necessary to create excess of quark-antiquark pairs. Not all the
quarks have the chance to form particles. Only those quarks experiencing
special conditions in presence of the colored field will form a particle
that satisfies the selection rules. Moreover, the strength of the color
field depends on the separation distance $r_{ij}$ . 

The number of parton pairs is $%
%TCIMACRO{\U{bd}}%
%BeginExpansion
{\frac12}%
%EndExpansion
N(N-1)$ and may be approximated as $%
%TCIMACRO{\U{bd}}%
%BeginExpansion
{\frac12}%
%EndExpansion
N$ for largevalues of $N$\textit{.}

$N$ depends mainly on the available energy required for creation of
quqrk-quqrk pairs. We follow the thermodynamic regime that starts by
calculating the partition function and The translational partition function
is then,

\bigskip 
\begin{equation}
Z_{tr}=\dint ...\dint \dprod\limits_{j}^{3N}\Psi _{j}^{\ast
}e^{-H_{tr}/KT}\Psi _{j}dq_{j}/h  \tag{2-5}
\end{equation}

For the sake of indistinguishability we divide \textbf{Eq. (2-5)} by $N!$
and write $Z_{tr}=Z_{p}Z_{q}$ for the momentum $p$ and spatial space $q$. As
usual, integration over $p$ we get

\bigskip 
\begin{equation}
Z_{p}=\frac{1}{N!}\left( \frac{2\pi MkT}{h^{2}}\right) ^{3N/2}\simeq \left( 
\frac{e}{N}\right) ^{N}\left( \frac{2\pi MkT}{h^{2}}\right) ^{3N/2} 
\tag{2-6}
\end{equation}

While the integration over $q$, is not simply $V^{N}$ because of the
presence of the potential energy $W_{ij}$. Which means that $W_{ij}$breaks
the factorizability of $Z$. 

\bigskip 
\begin{equation}
Z_{q}=\dint ...\dint \dprod\limits_{j}^{3N}\Psi _{j}^{\ast }e^{-\sum
W_{tr}/KT}\Psi _{j}dq_{j}/h  \tag{2-7}
\end{equation}

\bigskip Notice that for the confinement property of the potential then the
potential energy has trivial effect at small values of $r_{ij}$ so it is
possible to write \textbf{Eq.(2-7)} as

\bigskip 
\end{subequations}
\begin{equation}
Z_{q}=\dint ...\dint \dprod\limits_{j}^{3N}\Psi _{j}^{\ast }[1+Exp(-\sum
W_{ij}/kT)-1]\Psi _{j}dq_{j}  \tag{2-8}
\end{equation}

\bigskip \qquad \qquad \qquad \qquad \qquad \qquad \qquad \qquad \qquad
\qquad \qquad \qquad \qquad \qquad \qquad \qquad \qquad\ \ \ \ \ \ \ \ \ \ 
\begin{equation*}
=V^{N}+\dint ...\dint \dprod\limits_{j}^{3N}\Psi _{j}^{\ast }[1+Exp(-\sum
W_{ij}/kT)-1]\Psi _{j}dq_{j}
\end{equation*}

For the case where $W_{ij}/kT\ll 1$ we can expand the exponential in\textbf{%
\ Eq.(2-8)} as $Exp[W_{ij}/kT]-1\approx -W_{ij}/kT$ this is applied when $%
r_{ij}\prec 2$\ $r_{0}$ where $r_{0}$ is the quark radius ( $r_{0}=0.4fm$).
On the other hand, if $W_{ij}/kT$\ is not too small we can consider higher
order terms in the expansion of the exponential in \textbf{Eq. (2-8)}
.While, for $r_{ij}\succeq 2$\ $r_{0}$ the potential is increasingly with $r$
i.e. positively large so that  $Exp[W_{ij}/kT]\approx 0$ \ at extreme large
values of $r$, then,

\bigskip 
\begin{equation}
Z_{q}=V^{N}-\frac{1}{2}N^{2}V^{N-1}[4\pi \dint\limits_{2r_{0}}^{R}\Psi
_{1}^{\ast }[1+Exp(-W_{ij}/kT)-1]\Psi _{1}r_{1j}^{2}dr_{1j}  \tag{2-9}
\end{equation}

Now it is possible to calculate $F,S,U\ldots $ for this system in terms of $Z
$. \ $F=-NkT(\ln Z-\ln N+1)$. The entropy $S=-(\partial F/\partial T)_{N,V}$%
, then

\bigskip 
\begin{equation}
S=NkT\left( \frac{\partial \ln Z}{\partial T}\right) _{V}+Nk(\ln Z-\ln N+1) 
\tag{2-10}
\end{equation}

The internal energy $U=F+TS$; then

\bigskip 
\begin{equation}
U=NkT^{2}\left( \frac{\partial \ln Z}{\partial T}\right) _{V}  \tag{2-11}
\end{equation}

\bigskip The total internal energy $U$ is the important physical quantity in
the model and is directly related to the energy density and the particle
multiplicity production. The wave function $\Psi _{j}$ is assumed to be very
similar to that used by the parton model. Starting with the simplest
parametric form of the quark wave function,

\bigskip 
\begin{equation}
\widetilde{\Psi }_{a}(p)=Ce^{-\alpha p}\text{ \ \ \ \ \ \ \ \ }a\succ 0\text{
\ \ \ ,\ \ \ \ \ }p\succeq 0  \tag{2-12}
\end{equation}

p is the null momentum in the parton model  \textbf{[15],[16]}. The Fourier
transform of the quark wave function is then:

\bigskip 
\begin{equation}
\Psi (r)=\frac{C^{^{\prime }}}{(r+ia)^{2}}  \tag{2-13}
\end{equation}

\bigskip Where $C$ and $C\prime $ are normalization constants, while
\textquotedblleft $a$\textquotedblright\ is a fitting parameter. Again,
inserting \textbf{Eq. (2-13) }in\textbf{\ Eq. (2-5),} this gives the total
energy of the quark assembly of the nucleons. The range of the null momentum
p extends from zero up to Pmax. It is more convenient to express the wave
function and all other physical quantities in terms of the Bjorken scaling
variable with $x=P/P_{\max }$ lies in the range $\ 0\prec x\prec 1.$Particle
creation through formation and fragmentation of quarks is used through the
string model \textbf{\ [17]} consequently the multiplicity of particle
creation is calculated and compared with the recent results of Pb-Pb
collisions at incident momentum per nucleon of $40,80$ and $158GeV$.
(Experiment: \textbf{CERN-NA-049 (TPC))  [18-21]} 

\section{\protect\bigskip Results and Discussion}

We consider the case of central collisions in heavy ions of Pb-Pb. Let us
assume that overlap region at impact parameter b has a cylindrical form. The
number of nucleons and consequently the number of quarks Nn-part and Nq-part
for Pb+Pb collisions are calculated as \textbf{\ [22]:}

\bigskip 
\begin{equation}
N_{n-part}|AB=\int d^{2}S\text{ }T_{A}(\overrightarrow{S})\left[ 1-\left( 1-%
\frac{\sigma _{NN}^{inel}T_{B}(\overrightarrow{S}-\overrightarrow{b})}{B}%
\right) ^{B}\right]   \tag{3-1}
\end{equation}

\bigskip $\qquad \qquad \qquad \qquad \qquad +\int d^{2}ST_{B}(%
\overrightarrow{S})\left[ 1-\left( 1-\frac{\sigma _{NN}^{inel}T_{A}(%
\overrightarrow{S}-\overrightarrow{b})}{A}\right) ^{A}\right] $

\bigskip Where $\U{3c3} $ $_{NN}^{inel}$ is the inelastic nucleon-nucleon
cross section and 

$T(b)=\dint\limits_{-\infty }^{\infty }dz$ $\ n_{A}(\sqrt{B^{2}+Z^{2}})$ is
the thickness function. In calculating $N_{q-part}$; the density was
increased three times and $\sigma _{NN}^{inel}$ is re placed by $\sigma
_{qq}^{inel}$ We concentrate our calculation on the central collision region
where $b\preceq R_{P}+R_{T}\symbol{126}6fm$. On the average, the number of
nucleons (quarks) in the considered region is $250$ $(750)$. \textbf{Fig. (2)%
} illustrates the participant number of nucleons (quarks) according to
Glauber calculation \textbf{\ [22]}. 

The partition function for the $N$ particles is calculated according to 
\textbf{Eq. (2-9)} for separation distance $r_{ij}$ of the binary
interacting quarks. The family of curves of the partition function
represented in \textbf{Fig (3)} corresponds to temperature $20,40,60$ and $%
200MeV$. The temperature is very sensitive to the form of the nuclear
density. The temperature increases as the number of participant nucleons
from the projectile $N_{P}$ and the target nucleus $N_{T}$ come close $%
(N_{P}\longrightarrow N_{T})$. \ The temperature of the system is determined
according to the impact parameter and consequently depends on the number of
participating nucleons (quarks) \textbf{[1]}

\bigskip 
\begin{equation}
\zeta _{cm}=3T+m\frac{K_{1}(m/T)}{K_{2}(m;T)}  \tag{3-2}
\end{equation}

\bigskip 
\begin{equation}
\zeta _{cm}=[m^{2}+2\eta (1-\eta )mt_{i}]^{1/2}  \tag{3-2}
\end{equation}

In Eq. (3-2), $\zeta _{cm}$ represents the relativistic form of the center
of mass energy of a system of temperature $T$, (using system of units where
the Boltzmann constant $K=1$). The first term on the LHS,$(3T$ $or$ $3kT)$
is the thermal energy. The second term is the relativistic correction 
\textbf{[1]. }In Eq. (3-3) ) $\zeta _{cm}=[m^{2}+2\eta (1-\eta )mt_{i}]^{1/2}
$ describes also the center of mass energy for the particles in the overlap
region of projectile and target having nuclear densities $\U{3c1} _{p}$ and $%
\U{3c1} _{T}$ respectively with relative projectile density $\eta (r,b)$ at
position coordinate $r$ and impact parameter $b.$ \textbf{[1]} defined as

\begin{equation}
\eta (r,b)=\ \frac{\rho _{p}(r,b)}{\rho _{p}(r,b)+\rho _{I}(r,b)}  \tag{3-3}
\end{equation}

Solving the two equations\textbf{\ (3-2) }and \textbf{(3-3)}, it is possible
to find the temperature \textbf{T} at any $r$ and $b$. $t_{i}$ is the
incident kinetic energy per nucleon. $K_{1}$ and $K_{2}$ are the Macdonald
functions of first and second order. At low temperature $T\symbol{126}20MeV$
the partition function $Z(r,T)$ approaches flat behavior just above $%
r_{ij}\simeq 2$ $fm$. At higher temperature $(T=40,60$ $MeV)$ the $r_{ij}$\
dependence of $Z(r,T)$ becomes steeper. Approximate linear behavior is found
at high temperature $T=200$ $MeV$. 

On the other hand, the smooth variation of $Z(r,T)$ over all the range of
temperature $T,20\longrightarrow 200$ $MeV$ is given in\textbf{\ Fig. (4)}
with family of curves corresponding to $r_{ij}\sim 2,3,4,5$ and $6$ $fm.$
Steep drop of $Z(r,T)$ is observed in the cold region $(T<20MeV)$ for all
curves of  $r_{ij}\sim 2,3,4,5$ and $6$ $fm$. This is followed by smooth
increase toward the hot region up to $T\sim 200$ $MeV$. 

The changes of the partition function express the behavior of the quark
chemical potential $\mu $ inside the hadronic system through the well known
relation: $\mu =-kT\ln Z$

\textbf{Fig (5)} shows the change of quark chemical potential in a
temperature range up to $200$ $MeV$. The total energy $U(r)$ dissipated in
the nuclear interaction region due to the binary quark collisions is
represented in \textbf{Fig (6)} in the temperature range $T<200$ MeV. The
curves are plotted for quark-quark separation distances $r_{ij}\sim 2,3,4,5$
and $6$ $fm$. $U(r)$ has positive values in the small range $r_{ij}\sim 2,3,4
$ $fm$, where the quarks are approximately free and can carry enough energy
to create particles. However $U(r)$ has negative values in the large range $%
r_{ij}\sim 5$ and $6$ $fm$ where the quarks are mostly confined by the
string potential.

A Monte Carlo program is used to simulate the particle production in the
frame of String Model  \textbf{[17]}. The particle production is considered
as a tunneling process in a colored field. Because of the 3-gluon coupling,
the color flux lines will not spread out over the space as the
electromagnetic field lines do but rather be constrained to a thin tube like
region. Within this tube, new $q\overline{q}$ pairs can be created from the
available field energy. The original system breaks into smaller pieces,
until only ordinary hadrons remain. In the field behind the original
outgoing quark $q_{0}$  a new quark pair $q_{1}$ $\overline{q}_{1}$ is
produced so that the original one $q_{0}$ may join with a new one $\overline{%
q}_{1}$\ \ to form a hadron $q_{0}$ $\overline{q}_{1}$ \ leaving $\overline{q%
}_{1}$unpaired. The production of another pair $q_{2}$ $\overline{q}_{2}$\ \
will give a hadron $q_{1}$ $\overline{q}_{2}$ etc. From this assumption one
may find the resulting particle spectra in a jet. The possible meson
formation by the quark pairs are presented in \textbf{Table (1).} The
simulation process allows the production of all types of mesons with all
possible branching ratios taking into account the selection rules and the
conservation laws. 

In \textbf{Table (2)} the prediction of the simulation shows overall fair
agreement. In most cases at energies 20 and 30 A GeV the thermodynamic model
prediction exceeds the measured value by 11- 15\%. The prediction of the
model for pions and keons at energy 158 A GeV gives values little bit under
estimation with respect to the experimental values. However the calculated
value for the $\phi $ particle gives unexpected result (double the
experimental value) this may be due to the fact that the $\phi $ comes from
the channel of $s\overline{s}\ \ $quarks. In our model the production of u,
d and s pairs were considered equally probable; this seems in contrast to
the real case. 

Unfortunately, the Monte Carlo code used in our calculation was designed by
our research group since 1995 \textbf{[17]}. In this code the charged
particles (mesons $\pi ^{+}$; $\pi ^{-}$; $\pi ^{0}$; $K^{+};K^{-}$, etc.)
are created through string mechanism and the recombination of the specific
quarks, irrelevant of the mechanism of production whether due to the decay
of resonance particle ($K\ast $,\ldots ) or not. We are working right now to
develop this code, taking into consideration most of the recent information.

On the other hand, the recent STAR measurements on the production of various
strange hadrons (K$^{0}$s, phi, Lambda, Xi and Omega) in $\sqrt{S_{NN}}$ =
7.7 - 39 GeV Au+Au collisions show that strange hadron productions are
sensitive probes to the dynamics of the hot and dense matter created in
heavy-ion collisions. The extracted chemical and kinetic freeze-out
parameters with the thermal and blast wave models as a function of energy
and centrality were studied and discussed by Xianglei ZHU \textbf{[23-24]} 

We also believe that hadron production in general is a good probe to study
hadron formation mechanism in heavy ion collisions. At high transverse
momentum, pT, the hard processes, which can be calculated with perturbative
QCD, are expected to be the dominate mechanism for hadron productions. It
was observed at RHIC that, at high pT, the RCP (the ratio of scaled particle
yields in central collisions relative to peripheral collisions) of various
particles \textbf{[25]} indicates dramatic energy loss of the scattered
partons in the dense matter (jet quenching). RCP of hadrons have been
measured also at SPS \textbf{[26, 27]} as well, though the limited
statistics restricts the measurement at relatively lower p$_{T}$ (0.3
GeV/c). Measuring the nuclear modification factor in heavy ion collisions at
this energy range, one can potentially pin down the beam energy at which
interactions with the medium begin to affect hard partons \textbf{[28]}.

\section{\protect\bigskip Conclusive remarks}

\begin{itemize}
\item \qquad New particles need special environment to be produced during
the heavy ion collisions.

\item \qquad Multiple collisions among the quarks of the nuclear system
should produce large enough energy compared with the particle chemical
potential. The strong colored field plays the role of a catalyst parameter
necessary for particle production. 

\item \qquad The free available energy U(r) has positive values in the small
interaction distance where the quarks are approximately free and can carry
enough energy to create particles. However U(r) has negative values in the
large interaction distance where the quarks are mostly confined by the
string potential.

\item \qquad The string fragmentation and defragmentation is applied for the
production of the different types of newly produced particles. 

\item \qquad Theoretical attempts to understand the energy dependence of the
suppression were undertaken. Calculations were based on the Glauber-Gribov
model, in which the energy-momentum conservation was implemented into the
multiple soft parton re-scattering approach.

\item \qquad The temperature is very sensitive to the form of the nuclear
density. The temperature increases as the number of participant nucleons
from the projectile $N_{P}$ and the target nucleus $N_{T}$ come close. 

\item \qquad The temperature of the system is determined according to the
impact parameter and consequently depends on the number of participating
nucleons (quarks).

\item \qquad The thermodynamic model prediction exceeds the measured value
by percentage 11- 15\%

\item \qquad The quark-hadron phase transition will be studied in a
forthcoming article through temperature-quark chemical potential phase
diagram.
\end{itemize}

\bigskip {\Large Acknowledgment}

This paper was funded by Deanship of Scientific (DSR), King abdulaziz
University, Jeddah, under grant NO.(136-130-D1432). The authors, therefore,
acknowledge with thanks DSR technical and financial support. The authors
also would like to express deep thank to Prof. M.T. Ghoneim from Cairo
University for his help in improving the language and overall style of the
manuscript. 

\bigskip 

{\LARGE Figure Captions}

\textbf{Fig.(1)} 3-parameter quark-quark binary potential

\textbf{Fig.(2)} Number of participant nucleons (quarks) in the overlap
region as a function of the impact parameter b for the Pb-Pb collision

\textbf{Fig.(3)} The partition function Z(r,T) as a function of the
separation distance between the interacting quarks. Different curves belong
to temperature values of (20 (red), 40 (green), 60 (blue) and 200 (black)
MeV).

\textbf{Fig.(4)} The behavior of the partition function Z(r,T) in a
temperature range up to 200 MeV

\textbf{Fig.(5)} The change of quark chemical potential in a temperature
range up to 200 MeV

\textbf{Fig.(6)} The total potential energy formed inside the nuclear system
in the temperature range T \TEXTsymbol{<} 200 MeV. The curves are plotted
for quark separation distances r$_{ij}$ $\sim $ 2 (red),3(green),4(blue),5
(yellow) and 6 (black) $fm$.

\bigskip {\LARGE Table Captions}

\textbf{Table (1)} All possible $q\overline{q}$ pairs with their probable
meson type formation

\textbf{Table (2)} Table (2) Yields of particle production at Pb-Pb
collisions at 20 and 30 and 158 AGeV. The measured data are taken from ref
[20, 21] for pions and keons only. The possible measured values are compared
with the prediction of the present model.

{\LARGE References}

\textbf{[1] }\ \ Mohamed Tarek Hussein, Nabila Mohamed Hassan, Naglaa
El-Harby, Turk.J.Phys. 24(2000) 501;

-\qquad J. Gosset, H. H. Gutbrod, W. G. Meyer, A. M. Pokanger, A. Sandoval,
R. Stock and G. D. Westfall, Phys. Rev., C16, (1977) 629

-\qquad J. Gosset, J. I. Kapusta and G. D. Westafall,Phys. Rev., C18 (1978)
844

-\qquad W. D. Myers, Nucl. Phys., A 296, (1978) 177

\textbf{[2]}\qquad M. T. Hussein, N. M. Hassan, N. Elharby, APH N.S. Heavy
Ion Physics 13 (2001) 277

\textbf{[3]}\qquad Huichao Song, Steffen A Bass, Ulrich W Heinz, Tetsufumi
Hirano, Chun Shen, Phys.Rev.C83 (2011) 054910

\textbf{[4]}\qquad ALICE Collaboration, Phys.Lett.B696, (2011), 328 

\textbf{[5]}\qquad A.V. Nefediev, Yu. A.Simonov, Phys.Atom.Nucl.71 (2008)
171-179

\textbf{[6]}\qquad Jean-Paul Blaizot, J.Phys.G34 (2007) S243-252

\textbf{[7]}\qquad Qing-Guo Huang, Phys.Rev. D74 (2006) 063513

\textbf{[8]}\qquad M.N. Chernodub, F.V. Gubarev, Phys.Rev.D76 (2007) 016003

\textbf{[9]}\qquad S.N.Syritsyn, J.D.Bratt, M.F.Lin, H.B.Meyer, J.W.Negele,
A.V. Pochinsky, M.Procura, M. Engelhardt, Ph. Hagler, T.R. Hemmert, W.
Schroers, Phys.Rev.D81, (2010) 034507

\textbf{[10]}\qquad M. I. Gorenstein, M. Hauer, O. N. Moroz, Phys. Rev. C77,
(2008) 024911 

\textbf{[11]}\qquad Yuichi Mizutani, Tomohiro Inagaki, Prog.Theor.Phys.125
(2011) 933

\textbf{[12]}\qquad http://www.wolfram.com

\textbf{[13]}\qquad R. Sommer, Nucl. Phys. B411 (1994) 839

\textbf{[14]}\qquad O. Kaczmarek, F. Karsch, F. Zantow and P. Petreczky,
Phys. Rev. D 70, (2004) 074505

\textbf{[15]}\qquad M. T. Hussein, A. I. Saad; J. Mod. Phys., 2010, 1,
244-250

\textbf{[16]}\qquad Hussein, N. M. Hassan, and W. Elharbi, IJMPA Vol. 18,
No. 4 (2003) 673-683

\textbf{[17]}\qquad M.T. Hussein, A. Rabea, A. El-Naghy and N.M. Hassan;
Progress of Theoretical Physics, 93, 3 (1995), 585

\textbf{[18]}\qquad NA49 Collaboration (N. Davis et al.). Phys. Atom.Nucl.
75 (2012) 661.

\textbf{[19]}\qquad NA49 Collaboration (G.L. Melkumov (Dubna, JINR) et al.). 

\ \ \ \ \ \ \ \ \ \ \ \ Nucl.Phys.Proc.Suppl. 219-220 (2011) 102

\textbf{[20}]\qquad NA49 Collaboration, Phys.Rev.C77, (2008) 024903

\textbf{[21]}\qquad Francesco Becattini et al; Phys. Rev. C 85, (2012) 044921

\textbf{[22]}\qquad M.K. Hegab, M.T. Hussein and N.M. Hassan, Z. Physics A
336, (1990) 345

[\textbf{23]}\qquad Xianglei ZHU, Acta Physica Polonica B Proceedings
Supplement vol. 5 (2012) 213.

\textbf{[24]}\qquad Xianglei Zhu, for the STAR Collaboration, Nucl.Phys.A830
(2009) 845c-848c

\textbf{[25}]\qquad J. Rafelski and B. M%
%TCIMACRO{\U{a8}}%
%BeginExpansion
\"{}%
%EndExpansion
uller, Phys. Rev. Lett. 48, (1982)1066 

\textbf{[26]}\qquad B. I. Abelev et al., Phys. Rev. C 77, (2008) 044908

\textbf{[27]}\qquad J. Adams et al., Phys. Rev. Lett. 98, (2007) 062301

\textbf{[28]}\qquad X. Wang (STAR
Collaboration), J. Phys. G 35, (2008) 104074

\end{document}